RESEARCH ARTICLE

# How does fake news spread? Understanding pathways of disinformation spread through APIs

**Lynnette H. X. Ng**[1] | **Araz Taeihagh**[2,3]

[1]Institute for Software Research, Carnegie Mellon University, Pittsburgh, Pennsylvania, USA

[2]Lee Kuan Yew School of Public Policy, National University of Singapore, Singapore

[3]Centre for Trusted Internet and Community, National University of Singapore, Singapore

**Correspondence**
Araz Taeihagh, Lee Kuan Yew School of Public Policy, National University of Singapore, 469B Bukit Timah, Rd, Li Ka Shing Bldg, Level 2, #02-10, 259771, Singapore.
Email: spparaz@nus.edu.sg and araz.taeihagh@new.oxon.org

**Funding information**
National University of Singapore, Grant/Award Number: CTIC R-728-109-002-290

## Abstract

What are the pathways for spreading disinformation on social media platforms? This article addresses this question by collecting, categorising, and situating an extensive body of research on how application programming interfaces (APIs) provided by social media platforms facilitate the spread of disinformation. We first examine the landscape of official social media APIs, then perform quantitative research on the open-source code repositories GitHub and GitLab to understand the usage patterns of these APIs. By inspecting the code repositories, we classify developers' usage of the APIs as official and unofficial, and further develop a four-stage framework characterising pathways for spreading disinformation on social media platforms. We further highlight how the stages in the framework were activated during the 2016 US Presidential Elections, before providing policy recommendations for issues relating to access to APIs, algorithmic content, advertisements, and suggest rapid response to coordinate campaigns, development of collaborative, and participatory approaches as well as government stewardship in the regulation of social media platforms.

**KEYWORDS**

application programming interface, code repositories, disinformation, fake news, platforms, social media

# INTRODUCTION

"Fake news" is commonly used to refer to news that is false and that could mislead readers/viewers. Under the umbrella term "fake news," there are three common categories: "disinformation," "misinformation," and "malinformation" (Shu et al., 2020). These three categories are segregated in terms of their intent. "Disinformation" intends to deceive, and hence common techniques involve targeting profiles and fabricating content. "Misinformation" does not have malicious intent; examples include urban legends and genuinely false information. "Malinformation" has an intent to inflict personal harm: hate speech and harassment fall under this category.

In this article, we examine "disinformation," which has the goal of deceiving people. We study the use of code repositories to access Platforms through APIs for spreading disinformation on the platforms and examine how actors with malicious intent can utilise the platforms for their purposes. We used our findings to inform platforms and governments on pathways of disinformation spread and how to address the issues identified. We consider the following goals of an actor with an intent to spread disinformation in a network: (1) dissemination of a message across the network; (2) information amplification on desired topics; (3) planting/altering the views of large groups of users. As an "actor" operating tools and technologies to spread disinformation on social media, to achieve these goals, one needs to: (1) join an organic network of users, so the actor's message reaches real users; and (2) hide one's trace to avoid detection and suspicion, which would reduce the effectiveness of the message.

Fake news on social media platforms has become a contentious public issue as social media platforms offer third parties various digital tools and strategies to spread disinformation to achieve self-serving economic and political interests and distort and polarise public opinion. We study disinformation campaigns in the context of social media platforms. While social media platforms are revenue-generating businesses that promote user account and content creation, they have inevitably led to malicious actors spreading fake news. To achieve a successful campaign, an actor must perform a series of actions on the platform, some of which depend on others. A sequential combination of these actions characterises a pathway. This study on the pathways for spreading disinformation seeks to identify specific methods and stages of spreading disinformation. This will facilitate identifying new procedures to ensure the reliability and accuracy of disseminated information and increase the significant transparency of artificial intelligence- (AI-) driven data collection and algorithmic mechanisms for scenarios like online content recommendation. The study also profiles the risks and threats of AI-curated and generated content, such as a generative pre-trained transformer (GPT-3) (Brown et al., 2020) and generative adversarial networks (GANs) (Goodfellow et al., 2014). While revealing the ethical issues involved in curating and delivering online content, the study will help develop policy and legal responses to tackle online disinformation.

In the following sections, we define multiple pathways of disinformation within a framework. Application programming interfaces (APIs) can be used to obtain data from the platforms or inject information to the platforms.[1] We categorise social media APIs and seek to understand how APIs facilitate disinformation. We review the most investigated platforms concerning disinformation campaigns. We then collect further information from open-source code repositories from GitHub and GitLab to understand how developers use APIs. We then develop a framework regarding how an actor may spread disinformation on social media platforms. Finally, we investigate a case study using the framework developed in relation to the 2016 US Presidential Elections before providing recommendations for platforms and governments to address the issues relating to access to APIs, algorithmic content, advertisements, and suggest rapid response to coordinate campaigns, development of collaborative and participatory approaches as well as government stewardship in the regulation of social media platforms.

# LITERATURE REVIEW

## Preliminary results of the examination of the literature

Appendix A1 of the Supporting Information Materials along with Tables S1 and S2 highlight the details of the literature review methodology. Our examination reveals that the existing body of work remains segmented in its focus on particular technologies to spread disinformation on platforms (e.g., bots, APIs, tweet content). The studies focusing on bots analyse them predominantly in the context of spreading political disinformation. Empirical studies that analyse data from code repositories are mainly primarily concerned with assessing the success of already executed disinformation campaigns, such as by analysing the effects and patterns of information propagation in response to a limited set of actions and only focus on a single platform (e.g., Twitter) (Kollanyi, 2016; Shao et al., 2018; Zhou & Zafarani, 2018). While these studies are useful to understand how different technologies operate on digital platforms and their effect on the spread of disinformation, there is a lack of integration of these methods that reflect how developers and platforms spread disinformation using a combination of tools and the different ways that APIs facilitate these processes. To the best of our knowledge, no study has attempted to characterise the different actions deployed through APIs on different platforms to spread disinformation.

## Literature examining actions that facilitate the spread of disinformation

An emerging body of research examines disinformation and the tools facilitating their spread on digital platforms. These include studies that provide frameworks of the types of fake news being spread (Jr et al., 2018; Machado et al., 2019) and conduct in-depth analysis on the construction of messages and the credibility of their creators (Zhou & Zafarani, 2018). Many studies perform broad reviews on various channels (both digital and nondigital) to spread disinformation, such as examining how countries utilise television, social media, internet trolls, and bot networks to conduct political disinformation campaigns targeted at other states (Moore, 2019) and algorithmic recommender systems' roles in influencing user engagement and disinformation spread on social media (Valdez, 2019). Other works have examined the digital technologies available in the entire ecosystem of services, including both platforms and other service providers that enable targeted political campaigns on a massive scale. In particular, they analyse the tools used for collecting and analysing behavioural data and how digital advertising platforms profile and customise messages targeted at different audience segments (Ghosh et al., 2018).

Many scholars focus on the role of bots in spreading disinformation, with a predominant focus on Twitter, whereas there are limited studies that analyse APIs' role specifically in the spread of disinformation on other platforms. Studies have examined the origins of bots (Kollanyi, 2016), the pathways through which they function and which platforms they usually target (Assenmacher et al., 2020), and several studies have developed typologies of bots. Several typologies for Twitter bots have been produced, including characterisation of their inputs and sources, outputs, algorithms and intent and functions (Gorwa & Guilbeault, 2020; Lokot & Diakopoulos, 2016; Schlitzer, 2018), and these distinguish between bots that are used to increase the reach of a message and those that amplify a political narrative in a certain direction (Bastos & Mercea, 2018). Other studies analyse the availability of and mechanisms through which bot code is traded on the Dark Net to facilitate malicious uses (Frischlich et al., 2020). They focus on the digital infrastructure provided by APIs but do not analyse the different actions taken through these APIs to spread disinformation.

Various studies model and conduct empirical analysis of the diffusion of disinformation in response to specific actions and campaigns. For instance, Tambuscio et al. (2015), in the computing literature, have developed a diffusion model of the spread of disinformation. Shao et al. (2018) analysed the messages spread by Twitter bots in response to the 2016 US presidential campaign and election, while (Santini et al., 2020) conducted an empirical analysis of Twitter bots' behaviour in amplifying news media links to two Brazilian news sites that manipulate news media entities' online ratings and the relevance of news.

# METHODOLOGY—DATA COLLECTION AND ANALYSIS

## The landscape of social media APIs

An API is a "programming framework that implements a set of standard behaviours" (Puschmann & Ausserhofer, 2017). This article classifies APIs into two categories: (1) official APIs; and (2) unofficial APIs. Official APIs involve a developer having a platform-issued developer key or an authentication secret. Social media platforms control the use of official API keys to varying degrees. With these keys, developers can gain access to two sets of data: (1) data without restriction, which refers to data that users choose to share publicly; and (2) data restricted only to information about the developer's account (See Appendix A2 and Table S3).

Unofficial APIs include APIs meant for internal purposes that are used by third parties for unintended purposes. For example, a developer can examine how an official app on a device exchanges data with the platform's remote server and attempt to mimic that communication to develop new applications (Russell, 2019). Another type of unofficial API is code repositories that employ the web scraping method.[2] Some methods involve downloading the HyperText Markup Language (HTML) page, then parsing the page and extracting elements that match specific texts before executing actions on the HTML elements. The ease of employing these methods in the absence of official APIs affects the number of developers that will harness the platform for their agenda.

Analysing the number and type of unofficial APIs relative to the number of official APIs used by developers is potentially valuable to understand the primary channels used to spread disinformation, and how usage evolves and differs across platforms. In addition, it can provide key insights to understand the extent to which platform operators are aware of or should be accountable for how developers take advantage of these APIs to spread disinformation. We do this by examining social media APIs, and the types of actions registered users perform. The API documentation pages and literature documenting discontinued or undocumented APIs were also analysed.

## Open-source code repositories interacting with social media platforms

### Methodology for from open-source code repositories

To understand the potential uses of social media APIs, we look to open-source code repositories. In this study, our primary data sources are the public access code repositories, GitHub and GitLab. We investigate only open-source code platforms where the code is readily available, and that facilitate open collaboration and reference of codes via search terms. We thus miss out on code in private repositories and the Dark Web.

GitHub is the largest online code repository for shared computer code, with over 50 million users and 100 million repositories (Github, 2020b), and is the fastest-growing open-source platform (Kollanyi, 2016). While most of the GitHub projects are developed for timely retrieval of updates or auto-liking of a close friend's posts, the public accessibility of the repositories enables parties with a malicious intent to adapt the available code and construct their own versions easily. Furthermore, GitLab has an integrated continuous integration/continuous deployment pipeline, enabling the developed code to be instantly deployed for production. As a result, if the code is intended for a bot, the bot can perform its programmed tasks swiftly once the code is deployed through the pipeline. The GitHub/GitLab code repositories were sampled using the GitHub/GitLab developer search APIs to identify codebases that perform tasks related to the constructed pipeline. This sampling was done by searching each individual platform name together with the word "bot."

We then used the user search API to obtain a user item for each repository. This user item contains the profile of the user, such as the number of followers, the number of following, the number of repositories and the location (Github, 2020a). We extract the user's declared location from the user item. To map the extracted location to a country, we queried OpenStreetMap using its Nominatim search engine API, which searches through a list of addresses to find the closest matching address and its country. OpenStreetMap is a community-driven map built by enthusiast mappers and geographical professionals using aerial imagery and GPS devices (OpenStreetMap Contributors, 2017). It contains detailed street data of places around the world, even down to street-side cafes, that enthusiasts have manually entered.

### Data analysis of collected code repositories

Over the month of May 2020, we collected 69,372 code repositories. Over 40,000 repositories were collected that pushed content to social media platforms. Most code repositories cater for the platform Telegram, followed by Twitter, then Facebook and Reddit. The distribution of code repositories is shown in Figure S1. Appendix A3 in Supporting Information Material provides details of the data collected (Distribution of code repositories across social media platforms, Word cloud of descriptions from code repositories, Origin of data repositories collected, and Distribution of Programming Languages used in API Repositories, in Figures S2–S5, respectively). The number of repositories created per month increased exponentially from 2014 to 2018 before decreasing from 2018 to 2020. To supplement this observation, we manually searched social media API documentation on API changes. Sharp changes in the number of code repositories can be attributed to API changes of social media platforms (see Appendix A4 and Figure S6a,b).

## Official and unofficial API usages from code repositories

### Data collection of official and unofficial social media APIs

To understand how these code repositories perform their tasks, we sought to understand whether they use official or unofficial APIs. To this extent, we may infer how much official APIs provided by the platform facilitate the spread of disinformation. We first formulated a list of keywords relating to official and unofficial methods of accessing platform data and functionality. The keywords of official methods were collected through manual inspection of each social media platform's API documentation. The initial analysis of code repository content showed that repositories using official APIs typically contain an authentication string that uses the same name as is stated in the social media platforms' API documentation. Unofficial methods of accessing platform data and functionality typically comprise web scraping methods, many of which were originally developed for web user interface testing (e.g., selenium). We examined the code content of several repositories to profile unofficial methods of performing actions on platforms.

**TABLE 1** Search terms and content identifiers used for the data collection on API implementation

| Platform | Search terms | Type | Content identifier |
|---|---|---|---|
| Twitter | "twitter bot", "twitter scraper", "twitter crawler", "twitter posting" | Official | "consumer_secret" |
| Twitter | "twitter bot", "twitter scraper", "twitter crawler", "twitter posting" | Unofficial | "lxml", "selenium", "data-item-id" |

We queried GitHub/GitLab code hosting sites to search within the code content for the keywords through the blob search functionality to understand the distribution of methods used by repositories to perform specific actions on platforms.[3] Table 1 lists a sample of specific content identifiers used, along with the blob search functionality for searching through the content of the code repository, which is used to indicate the usage of official or unofficial APIs. For a complete table, refer to Table S4. After obtaining the code repositories through the blob search functionality, we carried out a systematic deduplication of each repository as some repositories mentioned the content identifier more than once and were hence counted more than once. We then queried the code hosting sites to identify the countries of the users that created the repositories.

We performed the search on four main platforms—Twitter, Facebook, Instagram, and Reddit—as these platforms draw the most repositories and are most widely used in disinformation campaigns. This was followed by exploring the characteristics of the repositories through a timeline of repository activity concerning the release or discontinuation of social media APIs.

# THE FRAMEWORK FOR UNDERSTANDING THE PATHWAYS OF DISINFORMATION SPREAD

## The pathways of disinformation spread framework

After building our knowledge base through a literature search and drawing on data from categorising social media APIs, and data collection on the open-source development landscape of APIs, we developed a theoretical framework of pathways that can be used to spread disinformation on platforms. As was highlighted under the literature review section, the studies conducted so far are segmented. They focus on particular technologies to spread disinformation on platforms (e.g., bots, APIs, tweet content), predominantly focus on a specific context such as spreading political disinformation, assessing the success of already executed disinformation campaigns, or are limited to a set of actions and only focus on a single platform. While we appreciate these scholarly works and they are useful to understand how different technologies operate on digital platforms and their effect on the spread of disinformation, there is a lack of integration of these methods that reflect how disinformation is spread using a combination of tools and the how APIs facilitate these processes in various ways. To the best of our knowledge, the theoretical framework presented in this article is the first that characterises the different actions deployed through APIs on different platforms to spread disinformation. This theoretical framework contributes directly to understanding pathways for content distribution and content collection on social media platforms and applies to different platforms. Drawing from research in social sciences and computer science, we identify four key stages for the spread of fake news: Network creation; Profiling; Content generation; and Information dissemination. We examine these stages in-depth, identify and group actions that can be performed in each stage, and present the relationship between these stages in the rest of this section. We further highlight how through these pathways, the goals of the actor(s) with intent to spread disinformation on social media platforms such as dissemination of messages across wide networks, information amplification and reiteration on desired topics, planting/altering the views of large groups of users, and using influential users to spread their own messages may be achieved. This theoretical framework on the spread of disinformation is visualised in Figure 1, and a full tabular breakdown can be found in Table S5.

An actor can perform a singular action on a social media platform, like following another created account. The actions are grouped into four stages, Stage 1: Network creation; Stage 2: Profiling; Stage 3: Content generation; and Stage 4: Information dissemination. A sequential combination of these actions in linear stages is a "pathway," which ends at a goal. An example of a pathway could be Stage 1: "Create user account"; Stage 2: "Attribute-based selection of audiences"; Stage 3: "Text generation"; and Stage 4: "Engage with users"; this ends at "Join a human network." The framework illustrates various pathways, which may be activated in a parallel fashion over time, to reduce time delays in steps and amplify the information dissemination effect. For example, creating user accounts takes time, but profiling users during an event need not wait. Further, actors occasionally return to a previous stage or substage during the activation of a pathway after they have understood their selected audiences better and decide to perform further actions to enhance their information dissemination to the target audience.

We profile two main types of accounts: (1) "pseudo-accounts," which are accounts created by actors, such as bot accounts, and (2) "user-accounts," which are accounts that real human users create. By extension, the usage of the term "pseudo posts" refers to posts that are generated by the "pseudo-accounts," and "user posts" refers to posts written by human users.

From examining the literature collected in Section 3.1, we identified different actions associated with the use of APIs to spread disinformation. Each action was distinguished based on the mechanism through which it is executed and the platform(s) it is executed on. We then associated each action with data gathered on API usage in open-source code repositories to further distinguish the actions based on how APIs (official and/or unofficial) are used to executing these actions.

A categorisation of the action "inflate retweet counts" could be described as follows. Through manual inspection of social media APIs, Twitter provides a mechanism to auto-like posts, and Instagram provides an API to auto-like all posts on a particular feed. Inspection of collected data on API usage indicates code repositories that use official and unofficial APIs to like posts with certain keywords on Twitter. For Instagram, we only found code repositories that use an unofficial API to auto-like and auto-follow a particular feed. This is likely because developers cannot pass the strict Instagram app review processes required to obtain official API access and thus must use unofficial means.

Next, actions exhibiting similar objectives and characteristics were grouped and characterised according to the type of actions employed. For instance, the actions of identifying users based on their follower account and identifying users that have posts associated with particular interests (and more specifically, keywords) are categorised as "attribute-based selection of audience," where user-accounts are selected based on whether they have (or do not have) certain attributes. We conducted the process of examining and categorising action types iteratively and with reference to the large base of scholarly literature collected and referred to in Section *Preliminary results of the examination of the literature*.

Lastly, we grouped the actions into four overarching categories representing different stages of curating and spreading disinformation on platforms. We presented a flowchart in Figure 1 that maps the possible pathways of disinformation dissemination on platforms that are facilitated by APIs: (1) network creation; (2) profiling; (3) content generation; and (4) information dissemination. We examined platforms commonly used in disinformation campaigns—Twitter, Facebook, Instagram, and Reddit. The Russian Internet Research Agency (RIRA) produced around 4234 Facebook posts, 5956 Instagram posts and 59,634 Twitter posts (Howard et al., 2019), spreading disinformation by creating false personas and imitating activist groups. Operation Secondary Infektion (Ben Nimmo et al., 2020) used Reddit, among other social media platforms, in the 2016 US elections and the 2017 German and Swedish elections.

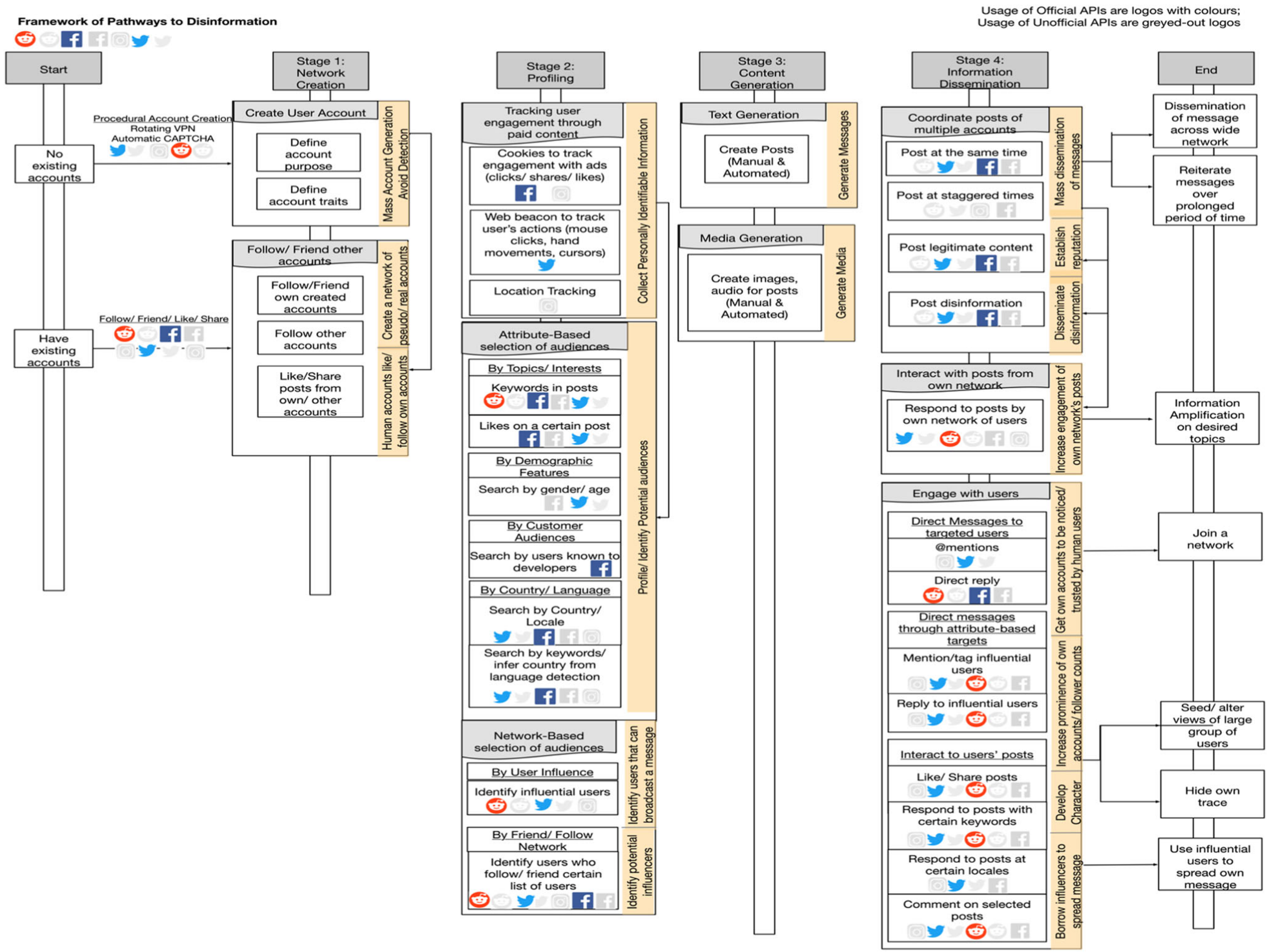

**FIGURE 1** The pathways of disinformation spread framework (note: there are two types of platform logos represented: coloured ones represent the use of official APIs, while greyed-out versions represent the use of unofficial APIs in achieving the objective). API, application programming interface

One key insight from the construction of the flowchart is that different social media platforms control the use of official API keys to varying degrees and hence play different roles in facilitating the spread of disinformation through different pathways and Table S5 summarises all the actions in the pathway to disinformation.

## Stage 1: Network creation

In the network creation stage, actors create a network of pseudo-accounts that will subsequently automate the execution of actions, each with a customised profile, identity, and purpose. Table 2 summarises the three main classes of actions that can be performed in this stage.

### Account creation

To begin the pathway, the actor needs to create a network of bots. Procedural account generation is the mass creation of individual accounts, where each account can have its own persona (i.e., age, gender, likes, dislikes). While official APIs do not provide this functionality, unofficial APIs, like web browser automation frameworks, allow for creating accounts (Jr et al., 2018). However, this is becoming increasingly difficult as social media platforms seek to reduce the creation of pseudo-accounts: Twitter requires a valid phone number, and Instagram requires solving a CAPTCHA. Some actors use a rotating Virtual Private Network to avoid detection by social media platforms, which avoids detection through rotating IP addresses.

Actors can also obtain existing accounts that are already active. This reduces their need to obtain the necessary verification details like phone numbers to create new accounts, and they thus inherit the accounts' purposes, traits, and network. It is possible to use a session cookie hijack attack to reset the account password (Ghasemisharif et al., 2018) or to exploit a flaw in the OAuth 2.0 security authentication protocol to obtain an access token (Hu et al., 2014). Actors can also obtain existing accounts through the Dark Net. Bots, or semiautomated accounts that mimic human behaviour, are readily found in underground markets in the Dark Net, which are forums and websites not indexed by search engines (Frischlich et al., 2020). The Dark Net requires a TOR browser, an anonymity-oriented browser, to access the websites. In general, the Dark Net provides bots across most social media platforms. Fake Facebook and Twitter accounts trade for around 5–9 Euros on average, and the highest price observed in a recent study was 42,435 Euros for a week's access to a botnet (Frischlich et al., 2020).

**TABLE 2** Summary of actions for Stage 1: Network creation

| Action type | Description | Facilitation of disinformation | References |
|---|---|---|---|
| 1a. Account creation<br>Create user account: define account purpose, define account traits | Use web browser automation to procedurally create account | Mass account generation, while avoiding detection through posing as real account | Jr et al. (2018) |
| 1b. Obtain accounts<br>Obtain existing Accounts: inherit account purpose, inherit account traits | Session cookie hijack attack | Obtain existing accounts/ network | Ghasemisharif et al. (2018) |
| | App impersonation using flaws in OAuth 2.0 security authentication protocol | | Hu et al. (2014) |
| | Obtain existing accounts/ bot network from Dark Net | | Frischlich et al. (2020) |
| 1b. Follow/friend own-created accounts | Follow other accounts | Create a network of pseudo-/ user-accounts | Woolley (2016) |
| | Like/share posts from own/ other accounts | Create a false impression of popularity | |
| | | User-accounts like/follow pseudo-accounts | |

### Follow/friend own/real accounts

Upon obtaining a series of online profiles, the actor then needs to create a network of pseudo-accounts by following or friending his own created pseudo-accounts. Political bots used the technique of following one's network of accounts to create a false impression of popularity across geographies (Woolley, 2016). He can then like/share posts from the pseudo-accounts to increase the attention given to the accounts since social media algorithms elevate more popular posts. Since the goal is to disseminate information to real users, the actor also needs to perform the same follow/friend actions on real accounts, hoping that a few will reciprocate. By liking/sharing posts from real accounts, the actor attempts to build trust with real users.

## Stage 2: Profiling

In the profiling stage, APIs are used to track user engagement with digital content on the platform or external websites, or track the location of the user's device, or track the different devices used by the user. Behavioural data collected here are used to extract knowledge about the user that will be used in the next stage to profile them and to tailor messages targeted to their unique preferences (Ghosh et al., 2018). With tools provided by Google, actors can insert web beacons into web pages that track users' actions in real-time. Individual profiles can be formulated for subsequent targeted information dissemination by studying user mouse clicks, hand movements, and hovering cursors. Some research is on building preference models given the preferences users exhibit through their online content (Recalde & Baeza-Yates, 2020) or the content they express likes for (Matuszewski & Walecka, 2020). An indication of these models' effectiveness is their ability to detect suicidal tendencies (Ramírez-Cifuentes et al., 2020), which by extension, allows the construction of models that can infer whether the user will believe and spread disinformation content. Table 3 summarises the three broad categories of actions under this stage.

### Tracking user engagement

To profile users' interests, actors may track their engagement with paid digital content on web pages. Using Google analytics, actors can create first-party web cookies to track clicks on advertisements and user search items and results. Actors can then directly collect user behavioural data to link these to an individual's personally identifiable information, like email addresses or mobile phone numbers. Facebook's Audience Network API reports user engagements for actors that construct their own Facebook pages, such as likes on posts and advertising campaigns' reach. While the API does not provide users' personal data, it is a tool that can be used to segment demographic groups so that actors can focus on individuals who are highly responsive to particular messages (Facebook, 2020a).

**TABLE 3** Summary of actions for Stage 2: Profiling

| Action type | Description | Facilitation of disinformation | References |
|---|---|---|---|
| 2a. Tracking user engagement | | | |
| Cookies to track engagement with ads/clicks/shares/likes/retweets | Insert cookies in webpage to track user preferences | Collect personally identifiable information | Ghosh et al. (2018) |
| Web beacon to track users' actions | Track user preferences by demographic groups with Facebook Audience Network API | | Facebook (2020a) |
| Location tracking | Integrate services like YouTube, Google Maps and Google Search into applications to leverage on the precise location mapping and rich consumer profiles that may be obtained through paid advertising services | | Russell (2019) |
| 2b. Attribute-based selection of audiences | | | |
| By topics/interests through… | Finding users that have certain keywords in posts or like specific posts | Profile/identify potential audiences for own narrative | Guilbeault (2018) |
| By demographic features through… | Identifying audiences through societal segments like gender/age | | Speicher et al. (2018) |
| By customer audiences through… | Using Facebook Pixel on webpage to collect data of visitors to the page for users that developers may have access to | | Facebook (2020b) |
| By country/language through… | Identifying potential targets through searching by country/locale, or keywords in target language | | |
| 2c. Network-based selection of audiences | | | |
| Identify influential users | Find users that have a large network structure | Identify users that can broadcast a message | Ghosh et al. (2018) |
| By friend/follow network | Identify users who friend/follow a predefined list of users of interest and select active users through network structure | Identify potential influencers | Speicher et al. (2018) |

## Attribute-based selection of audiences

Various actions can be taken to select the desired group of users based on whether they exhibit (or do not exhibit) a certain attribute (Guilbeault, 2018). This method draws on digital marketing and advertising ideas, including identifying users through societal segments, such as topics of interest inferred through their posts and likes, demographic features of gender and age, and geo-location factors such as country and language (Speicher et al., 2018). Most social media APIs provide a search functionality where actors can specify these demographic features as input parameters. Where official APIs are insufficient, code developers use unofficial API methods by mimicking a real human user's search in the search box, scraping the webpage, then performing algorithms on the text to identify texts of interest, and by extension, users of interest.

Another way to select the desired group of users is to search by customer audiences or users known to the actor. Facebook provides web page administrators with a Facebook Pixel, a $1 \times 1$ image that can be inserted on a web page to collect visitors' data (Facebook, 2020b). The web developer can then target Facebook users who have visited the webpage.

### Network-based selection of audiences

User-accounts can be selected based on which network they belong to, selecting for the network structure rather than the user attribute. Both official and unofficial APIs from several social media platforms allow actors to grab network information about users (i.e., their following, follower, and friend network). Using network analysis, actors may identify influential users with a large network structure to target as potential influencers to broadcast their message (Ghosh et al., 2018). At the same time, unofficial APIs may achieve the same functionality by performing web scraping on the following/follower/friend tab.

Another way of identifying potential influencers is through identifying users who follow/friend certain lists of users. The original list of users includes those that interest the actor: for example, a famous politician. Through the network structure, actors may identify other active users who can influence views.

## Stage 3: Content generation

After profiling social media users, it is then natural to wish to push content to the users. Before disseminating information, an actor must generate content. We define Stage 3 as Content Generation, which covers two main content generation techniques that may be used in this stage. The techniques used in this stage, and references, are summarised in Table 4.

**TABLE 4** Summary of actions for Stage 3: Content generation

| Action type | Description | Facilitation of disinformation | References |
|---|---|---|---|
| 3a. Text generation | Generate specific synthetic propaganda reflecting an ideology by fine-tuning language models | Generate messages for dissemination | OpenAI, 2019; Solaiman et al., 2019 |
| 3b. Media generation | Generate synthetic instances of audio/visual data that is very similar to real data | Generate profile images for procedural account generation, generate image posts for dissemination on the network | 4chan (2020) |

In this stage, actors generate fake content to push onto the social media platforms. Using such systems, actors can procedurally generate massive amounts of uniquely different content suited to the accounts' purposes. Unique content is less likely to get flagged by the platforms' bot algorithms and is more likely to stay undetected by the platform. These accounts can thus navigate the platform landscape along the disinformation pathways. Twitter user @DeepDrumpf was a bot that produces tweets based on a neural network language model relating to Donald Trump's tweets before his permanent account suspension.

The latest advance in text generation is the GPT-3 (Brown et al., 2020) by OpenAI. It is a massive language model, with 175 billion training parameters, trained on Wikipedia texts, digitised books and web links. The estimate of the amount of input data from Wikipedia texts alone is approximately 6 million articles. In contrast, its predecessor, GPT-2, used less information and 1.5 billion training parameters. The latest development has even proven the ability to respond to the essay “GPT-3 and General Intelligence” by David Chalmers, replying with a philosophical letter (Chatfield, 2020). In fact, in releasing its predecessor GPT-2, OpenAI, and Middlebury Institute of International Studies' Center on Terrorism, Extremism, and Counterterrorism (CTEC) has released a blog post stating its algorithm can be fine-tuned for misuse. By fine-tuning the models on ideologies, the models can generate specific synthetic propaganda (OpenAI, 2019; Solaiman et al., 2019).

Media generation typically uses GANs, an algorithmic system of pitting two neural networks against each other to generate synthetic instances of data similar to real data (Goodfellow et al., 2014). Actors may use these techniques to generate synthetic images and audio to support the information dissemination.

Occasionally, this stage may support *Stage 1: Network creation*, where actors generate profile images, which enhances creating a network of profiles that do not use real identities and that are not similar to each other. A thread in 4chan (2020) elaborates on possible sites actors can use to generate user selfies so the account cannot be traced to a real identity. This thread, titled “Fake Twitter Account Support General,” discusses and “shares tips on creating and maintaining wolf in sheep's clothing Twitter accounts.” It also promotes the site https://thispersondoesnotexist.com as a great site for user selfies because “It uses AI to randomly generate faces, so your account can't get traced back to a real person who isn't you.”

## Stage 4: Information dissemination

Once users are profiled, in Stage 2, targeted messages can be constructed to appeal to the users' interests. The construction of targeted messages was described in the discussion of *Stage 3: Content generation* above. Upon constructing targeted messages, actions can be taken to disseminate information on a large scale (see Table 5 for the summary).

### Coordinating posts of multiple pseudo-accounts

When multiple pseudo-accounts post simultaneously or at staggered intervals, the actor achieves mass dissemination of a message across a wide network. However, it is a prerequisite that this network interacts with real accounts to influence human users. The pseudo-account network may first post legitimate content to establish a reputation (Linvill et al., 2019), gaining real accounts as followers before posting and disseminating disinformation. Studies characterising social bot networks in the 2017 French presidential election support discovery of a coordinated behaviour of bots that have been identified through machine learning algorithms (Ferrara, 2017).

**TABLE 5** Summary of actions for Stage 4: Information dissemination

| Action | Description | Facilitation of disinformation | References |
|---|---|---|---|
| 4a. Coordinate posts of multiple accounts | | | |
| Posting timing | Post at the same time | Mass dissemination of messages across wide network | Ferrara (2017); Ghosh et al. (2018); Shao et al. (2018) |
| | Post at staggered intervals | Reiterate messages over a prolonged period | Neudert (2017) |
| Post legitimate content | Relay breaking news, post about a mundane user's daily life | Establish reputation | Bastos and Mercea (2018); US Senate, (2020) |
| Post disinformation | Construct messages relating to desired narrative and post in social network | Disseminate disinformation using multiple accounts to amplify desired topics | Linvill and Warren (2020) |
| 4b. Interact with posts from own network | | | |
| Respond to posts by own network of users | Use own botnet to like own posts, as platforms' recommendation algorithms tend to prioritise content with more responses | Increase engagement of own network's posts, leading to information amplification | Bastos and Mercea (2018) |
| 4c. Engage with users | | | |
| Direct messages to targeted users | Send messages to users via @mentions or direct replies | Get pseudo-accounts to be noticed and trusted by human user-accounts | Chamberlan (2010); Linvill and Warren (2020) |
| Direct messages through attribute-based targets | Gain attention from attribute-based targets through mentioning/tagging influential users, or replying to posts by influential users | Increase prominence of pseudo-accounts<br>Plant/alter views of large group of users<br>Use influential users to spread own message, hiding trace | Lokot and Diakopoulos (2016) |
| 4d. Interact with users' posts | | | |
| Like/share posts | Promote posts of network of users | Develop character | Lokot and Diakopoulos (2016); Zannettou et al. (2019) |
| Respond to posts at certain locales<br>Comment on selected posts | Gain attention from attribute-based targets through responding with replies, comments, likes or shares | | |

## Interacting with posts from own network

Responding to posts through likes, shares and comments to the actor's own network of pseudo-accounts increases engagement with the particular posts (Zannettou et al., 2019). As social media platforms prioritise posts through their interaction count, this action amplifies the information the posts seek to disseminate and propagate the message through the platform.

### Engaging with users

Interacting with user posts through likes, retweets/shares, replies/comments is done to develop the pseudo-account's character and to increase the prominence of the pseudo-account. The Internet Research Agency generally uses replies/comments, rather than retweets/shares earlier on in their campaigns, to build the character of the pseudo-account and to place themselves in the network (Linvill & Warren, 2020). Direct messages to targeted users get the actor's account noticed by other user-accounts (Chamberlain, 2010). Pseudo-accounts can send direct messages to attribute-based targets, such as influential users identified in the previous stage through network profiling. This action seeks to use influential users to spread a message through their already-established network, which they have a reputation in, and to hide any trace of the pseudo-accounts, such that the origin of the message is unclear. Interacting with user posts via likes/shares and comments on selected posts seeks to increase the prominence of pseudo-accounts. Based on accounts obtained from attribute-based targeting in the previous stage, the pseudo-accounts may respond to posts with certain keywords or locales, hence building pseudo-character, and increasing the prominence of the pseudo-account (Lokot & Diakopoulos, 2016).

For example, Twitter's API facilitates Stage 2 via providing an API that streams tweets containing the requested keywords and also facilitates Stage 3 by providing an API that allows replying to a tweet. An unofficial API method would involve grabbing tweet IDs via web scraping methods, then replying to the tweet by firing up a Chrome web browser framework to perform the necessary mouse clicks and keyboard actions. To supplement the construction of the reply, tools from the Supplementary stage: Content generation may be used to construct a tweet reply automatically.

# DISCUSSION

## Application of the framework

Using the developed framework, we seek to understand how the actions facilitated by APIs are implemented in practice to spread disinformation. The same action can differ in its intended objectives. For example, having multiple profiles post simultaneously can meet the objectives of mass dissemination of a message, reiterating messages across a prolonged period, or amplifying information on a desired topic. Also, a particular combination and sequence of actions is required to meet a certain objective most effectively. Hence, it is valuable to identify and examine the different pathways through which actions are taken to meet different goals for disseminating disinformation. This section examines a case study of the 2016 US Presidential Election and applies the developed typology to demonstrate how the actors constructed their disinformation pathway. Figure 2 provides a pictorial overview of the portions of the framework of disinformation that were activated during this event, and Table 6 summarises the actions and operations at each of the steps of the framework.

For the activation of Stage 1, the RIRA constructed a large network of pseudo-accounts. The technique of creation of numerous accounts increases the reach of a disinformation campaign when launched at Stage 4. Typically, as mentioned in Stage 1, each account has its own personality and type of content it posts. The agency created massive amounts of pseudo-accounts on Twitter from 2015 to 2017 to sow discord during and after the 2016 US Presidential Election. This agency coordinated a complex multiplatform disinformation campaign through Facebook, Twitter, Instagram and Reddit (Lukito, 2020). There were also Twitter pseudo-accounts comprising accounts that previously had human operators and automated bots (Badawy et al., 2018). While we cannot identify the percentage of these accounts, they were active in the months before the election and pushed content reflecting both right-wing and left-wing ideologies, news feeds, and hashtag gaming content (Linvill et al., 2019). Peaks in account creation times on Twitter and Reddit corresponded to the lead-up to the announcement of the Republican nomination (Zannettou et al., 2019). To avoid detection, profile descriptions of pseudo-accounts were integrated into the target community by posing as news outlets or journalists and US persons affiliated with political parties (Linvill & Warren, 2020; Mueller, 2019).

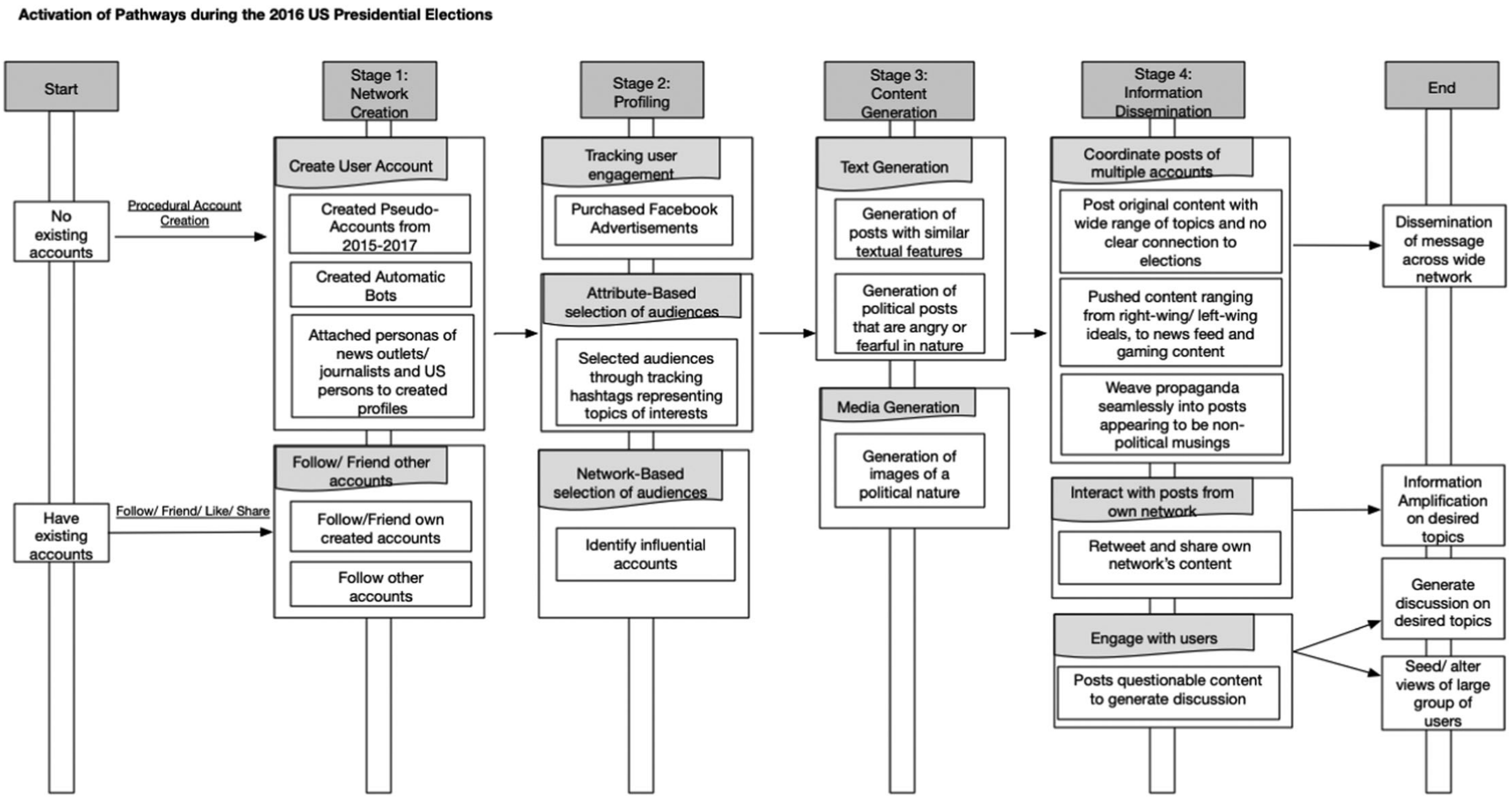

**FIGURE 2** Activation of the pathways of disinformation spread in the 2016 US Presidential elections

**TABLE 6** Activation of framework of disinformation spread during the 2016 US Presidential Election

| Description | Activation during the 2016 US Presidential Elections | References |
| --- | --- | --- |
| Stage 1: Network creation | | |
| Procedural account generation/create user account | Created pseudo-accounts on Facebook, Twitter, Instagram and Reddit, Tumblr, YouTube between 2015 and 2017 | Lukito (2020); Mueller (2019) |
| | Created automated bots | Badawy et al. (2018) |
| | Attached personas of pseudo-accounts of news outlets/journalists and US persons affiliated with political parties to created profiles | Linvill and Warren (2020); Mueller (2019) |
| Follow/friend own network/ other user-accounts | Increased their follower count by following/friending own/other networks, especially the networks targeted for infiltration | Linvill and Warren (2020) |
| Stage 2: Profiling | | |
| Tracking user engagement | Purchased Facebook advertisements | Mueller (2019) |
| Attribute-based selection of audiences | Selected audiences through tracking hashtags (topics of interests) | Zannettou et al. (2019) |
| Network-based selection of audiences | Identified influential accounts | Linvill and Warren (2020) |
| Stage 3: Content generation | | |
| Text generation | Generation of posts with similar textual features | Badawy et al. (2019) |
| | Generation of political posts that are angry or fearful in nature | Miller (2019) |
| Image generation | Generation of images of a political nature | Zannettou et al. (2020) |
| Stage 4: Information dissemination | | |
| Coordinate posts of multiple accounts | Posted original content with a wide range of topics and with no clear or overt connection to the elections | Linvill et al. (2019) |
| | Pushed content ranging from right-wing and left-wing ideals, to news feeds and hashtag gaming content | US Senate (2020) |
| | Weave propaganda seamlessly into posts appearing to be nonpolitical musings of an everyday person | |
| Interact with posts from own network | Retweeted and shared their own network's content | Zannettou et al. (2019) |
| Engage with users | Posted information of questionable content on Reddit to generate discussion | Zannettou et al. (2019) |

In Stage 2, the Internet Research Agency's pseudo-accounts used a network-based selection of audiences to identify influential accounts (Linvill & Warren, 2020). They selected audiences through topics of interest, by filtering for hashtags, and by users' self-reported countries (Zannettou et al., 2019). They also purchased Facebook advertisements to advertise their political ideals, possibly to obtain a user base that could be used for further targeting (Mueller, 2019). Through audience selection, populations that are likely to believe certain narratives and spread them can be selected.

Pseudo-accounts use Stage 3, content generation, to create content for their accounts that spread the disinformation narratives they wish to propagate, sometimes with the help of content generation tools. While there has been no deterministic work on the percentage of generated text deployed by the pseudo-accounts, many scholars have characterised bot-generated text through content features like length and entropy of texts (Badawy et al., 2019). For example, text content generated on Twitter include policy-related topics using angry and fearful language (Miller, 2019). To increase the engagement and credibility of their posts, these accounts regularly share images, mostly of a political nature (Zannettou et al., 2020).

In the initial phase of the activation of Stage 4 of the Internet Research Agency campaign, the pseudo-accounts were primarily producing original content before shifting to a mix of half original and half external retweets, then to constant tweets. The pseudo-accounts further engaged with the user accounts of the networks they wished to join through replies. At the same time, they posted original content to establish the reputations of the pseudo-accounts. These tweets had a wide range of topics and had no clear or overt connection to the elections (Linvill et al., 2019). In retweeting and sharing their own network's content, they created retweets and network content, often specific to the source of the pseudo-accounts (Zannettou et al., 2019). This erratic behaviour of tweeting allows the pseudo-accounts to hide under the radar and remain unsuspected both by the Twitter population they are targeting and Twitter's detection algorithms.

The application of the framework does not always follow a sequential order and, at times, can be more complex. Actors may move between stages, occasionally returning to previous stages to serve their aims. Studies show that regardless of which stage the actors are in, they occasionally return to Stage 1 to increase their follower counts and expand their network by following other accounts. Following this, when the influence network has grown large enough, agents jump to Stage 4 and perform information dissemination. This is usually the campaign's final information push, where pseudo-accounts amplified their content through retweets on Twitter (Linvill & Warren, 2020) and post questionable content to generate discussion on Reddit (Zannettou et al., 2019).

## Policy recommendations

Based on the research conducted in this study on examining how code repositories can be used to access platforms through APIs for spreading fake news on the platforms, the following policy recommendations are proposed for platforms and governments in the governance of social media APIs. In the short term, we recommend platforms monitor Stage 1: Account Creation and Stage 3: Content Generation, and both platforms and governments to monitor Stage 4: Information Dissemination. In addition, governments can adopt a multicollaboration approach to addressing medium and long-term disinformation spread.

## Proactive monitoring of the account creation process and access to the APIs

To monitor Stage 1: Account Creation, platforms need to proactively balance their efforts to increase the number of active accounts with their ability to monitor the users to identify accounts and users with malicious intentions. Most platforms currently require phone authorisation during the creation of accounts, limiting the number of accounts a person can create. However, actors can bypass this restriction by purchasing multiple phone cards. Restricting platform usage to only one account per user will not work. Many users will want to have multiple accounts for different uses, for example, an account for posting news and another account for posting photographs. Platforms should require description of the account's purpose, and perform checks on the account content, to see if the account is exhibiting behaviour on the platform in line with their declared purpose. In case of divergence from the intended purpose, appropriate action against the accounts that violate community guidelines or the terms of service should be taken. These checks and punitive actions, while can be to some degree automated, should require a human-in-the-loop to allow for discretion as users' tastes, hobbies, and behaviours can evolve in a complex manner and should have a transparent appeal mechanism that is human-driven.

However, we acknowledge that platforms are ultimately revenue-generating businesses that depend on user signups and content generation. Unfortunately, their business of gathering users and promoting content not only generates revenue but also facilitates the manipulation of content in its current form. While previous incidents such as the 2016 US Presidential Elections have forced the platforms to put more safeguards in place, heavy monitoring and gatekeeping may run counter to the platform's business models.

While platforms should maintain a free API access that provides restricted amounts of data and premium tiers that provide a larger volume of data to businesses or facilitate scholarly research in examining online platforms, they need to exercise further scrutiny in granting API access. They should require a statement of purpose from the API users before granting access to evaluate their intentions. Platforms should check in with the users and require output after a few months, both in the form of software applications and/or prototypes and research papers, to prevent abuse of APIs. For instance, the Twitter Academic API is restricted to academic institutions with a clearly defined project and requires applicants to provide potential academic publication venues as verification (Twitter, 2020). Such initiatives can be further extended across all social media platforms to businesses, academics, or even to hobbyists and other nonacademic researchers, as long as they have a defined project and visible results/outputs in a reasonable amount of time, depending on the size of their projects.

## Proactive monitoring of algorithmic content

For Stage 3: Content Generation, besides monitoring user behaviour, platforms should monitor content. Algorithmic content like generated texts and images pose a problem as these tools provide the means for content to be quickly generated to spread messages. While some academic work has been done in detecting generated texts (Gehrmann et al., 2019), much of it is still in its infancy. Platforms should work closely with academic researchers that develop generation tools to develop corresponding detection tools. For verified research groups, platforms can provide anonymized data to develop detection algorithms, then further integrate the resulting detection models into their platform monitoring system.

## Monitoring suspicious activities and rapid response to coordinated campaigns

For Stage 4: Information Dissemination, beyond examining individual user's behaviour on the platform, there should be a concerted effort to monitor activities of groups of users. These activities are usually carried out by a network of bots and can be in the form of coordinated campaigns. To develop a rapid response to coordinated campaigns, Twitter has built a Trust and Safety team to monitor such activities and is committed to removing coordinated campaigns with a zero-tolerance policy (Twitter, 2021). However, during the COVID-19 pandemic, despite Twitter claiming there is no significant coordinated attempt happening on the platform, academic researchers have observed that politically motivated coordinated inauthentic activity resulted in 5 million impressions on conspiracy theory tweets (Graham et al., 2020). Platforms should also work closely with researchers to leverage their insights on how to monitor and detect suspicious activities and coordinated campaigns rapidly.

## Supervision of the advertisements

Actors may use advertisements in Stage 3: Content Generation as well to disseminate information. Advertisements are a valuable source of revenue for social media platforms. Platforms should verify the legitimacy of the advertising entities and continuously monitor their content to ensure proper use of their advertising channels. Also, similar to Facebook's "report a spam account" feature (Facebook, 2020d), there should be a "report spam/inappropriate advertisement" feature that leverages crowd-sourced information to verify the advertising content matches the advertiser's profile.

## Restriction of online information dissemination during major events

Governments also can act in Stage 4: Information Dissemination. In an event-based governance strategy, governments can design restrictions on online campaigning. Social media has become a popular platform for online campaigning, with many politicians campaigning online and many political watchers expressing their views through these platforms (Margetts, 2017). One example is the 2020 Singapore General Elections, which implemented a "cooling-off day," where online campaigning was not allowed the day before the polling day. Thus, any information passed through social media during that day can be classified as unofficial information, or in the malicious form, disinformation. This regulation can be further extended to major events by restricting the dissemination of information through online media, for example, for the President's inauguration. Further, educating the public through the official channels on online platforms regularly for information dissemination, engaging the crowds and creating crowd capital over a period of time (Prpić et al., 2015; Taeihagh, 2017) reduces the chance the public will resort to unreliable sources of information.

## Government stewardship in regulating social media

Governments need to exercise stewardship in balancing the risks of innovation with regulations when it comes to disruptive technologies (Tan & Taeihagh, 2021). In the context of social media, balancing the risks of privacy and the consequences of disinformation is the key. The Singapore government has enacted Protection from Online Falsehoods and Manipulation Act (POFMA) to fight disinformation. While the UK government has not enacted any laws yet, it has performed a Cairncross Review to make recommendations to place online platforms under regulatory supervision (Feikert-Ahalt, 2019). The Singapore approach has proved to be effective as the law has been invoked several times on Facebook pages making false claims, and most notably, has led to the decision of Google to stop running political advertisements in Singapore (Meyer, 2020). This law relies on the crowd to point out and prevent disinformation spread and has gained widespread acceptance from the public, who turned “POFMA” into an everyday slang with “POFMA you” (Ng & Loke, 2020), to indicate the possibility the recipient may be disseminating fake information.

## Collaborative and participatory approaches in addressing the medium-term and long-term disinformation spread

The medium-term and long-term spread of disinformation across social media platforms remain black boxes. While in the near-term, governments can enact regulations around events and partner with fact-checking websites to clarify information, strategies must be designed to address longer-term disinformation spread. Combatting the spread of disinformation also requires identifying disinformation and informing the public of the fake information. Governments alone cannot address the spread of disinformation nor identify all the disinformation occurring during a certain event. Governments and platforms should work together to fact-check content and should further collaborate with Fact-Checking Websites such as Poynter, which hosts the International Fact-Checking Network (IFCN), a series of chapters worldwide that support fact-checking by volunteers. These networks have amassed large resources and have also been actively fact-checking the COVID-19 events and US elections. Facebook has taken a step in this direction by working with the IFCN to identify fake news and review content, and the platform uses this information to inform users about the validity of the news (Facebook, 2020c). There have been accusations against one of IFCN's verified signatories of incorrect fact-checking (Poynter, 2019), highlighting the importance of a multipronged fact-checking approach across several institutes and organisations for transparency.

In addition, disinformation identification can be crowd-sourced. Citizens can be more engaged in a participatory approach, such as reporting or clarifying information through public chat channels. When crowdsourcing platforms are properly designed, the effectiveness of the participatory platform is enhanced (Liu, 2021). One attempt is the SureANot WhatsApp group developed by undergraduate students in Singapore for people to provide tip-offs to information they are unsure if it is real or fake (Lee, 2020). Platforms and governments should tap on and promote these groups to take advantage of the heterogeneous knowledge of individuals through crowdsourcing (Prpić et al., 2015) for fact-checking efforts to help more people identify disinformation.

# CONCLUSION

This article has examined social media APIs and highlighted the various types of actions code developers may use to harness these APIs to push or pull content to/from social media platforms. Through a data analysis of these code repositories, we observe the following trends: (1) the number of repositories increased exponentially from 2014 to 2018, before decreasing from 2018 to 2020; (2) the change in the number of repositories created coincided with dates when platform API terms and conditions changed; (3) most code repositories are constructions of bots for the platforms.

Afterwards, by further searching the code base of the code repositories relating to social media APIs, we characterised the mechanisms code repositories use to interact with social media platforms using official or unofficial APIs. We observed that most code repositories use official platforms APIs, and the most preferred programming language is Python. This systematic analysis provides an understanding of the interaction between the social media API and the code development landscape.

By characterising sequences of these pathways, we constructed a nonlinear typology, organising the pathways for content distribution and content collection on social media platforms into a systematic framework. The framework consists of four main stages: network creation, profiling, content generation and information dissemination. Through this framework, some of the goals that actors with an intent to spread disinformation on social media platforms may achieve include: dissemination of messages across wide networks, information amplification and reiteration on desired topics, planting/altering the views of large groups of users, and using influential users to spread their own messages.

We finally demonstrated the application of the Pathways of Disinformation Spread Framework through the case study of the 2016 US Presidential Elections and illustrated the key stages of the framework that were activated during this event. This was a singular application of the theoretical framework established in this contribution. The framework can inform future empirical research to differentiate the techniques actors use across different events, regions, and platforms through in-depth single and comparative case studies. Furthermore, the framework can be further expanded to examine pathways for the spread of disinformation and actions within each stage of the framework, which was primarily on the spread of information, to include the use of AI and different types of AI-generated content and examine practical challenges that arise from to counter AI use for the spread of fake news. We further proposed policy recommendations platforms and governments can adopt to regulate the digital landscape and create a healthy online information environment.

Online misinformation has important ramifications not only for individuals but also for society as a whole. Disinformation can be a threat to the society where the spreading of false narratives could lead to negative health outcomes as demonstrated through the COVID-19 pandemic through leading citizens to make decisions that are harmful to their health, such as avoiding seeking medical treatment or vaccination, or result in anger, distrust and offline violence in the society through protests and contribute to events such as the 2021 United States Capitol attack, which could threaten society's peace. Widespread disinformation makes governance difficult due to the lack of trust of citizens in the information they receive. In the digital world, citizens mainly consume information through online sources, notably social media. We hope that this study facilitates a more nuanced examination of the spread of disinformation on social media platforms through a framework characterised by the actual usage of platform APIs by code developers. We further hope that the findings of this study are useful to both the research and policy practitioner communities and allow them to glean a better understanding of the pathways by which fake news may be spread on social media platforms to further formulate policies to identify and address emerging disinformation.

## ORCID

*Araz Taeihagh* http://orcid.org/0000-0002-4812-4745

## ENDNOTES

[1]See Appendix A2 in the Supporting Information Material for more information about APIs.

[2]Web scraping refers to downloading a website and extracting data from it. Web scraping can be done manually through accessing the web through HTML protocol using a web browser or can be automated using bots or crawlers.

[3]The blob search functionality allows us to search keywords through the code contents and return the code snippets that contain the keywords of interest. Blobs are often collections of audio, images, or executable code objects that are stored as binary data in a single entity.

## SUPPORTING INFORMATION

Additional supporting information may be found in the online version of the article at the publisher's website.